\documentclass[%
 reprint,
 showkeys,
nofootinbib,
nobibnotes,
 amsmath,amssymb,
 aps,
floatfix,
]{revtex4-2}

\usepackage{graphicx}
\usepackage{dcolumn}
\usepackage{bm}
\DeclareMathOperator*{\argmin}{arg\,min}

\begin{document}

\preprint{APS/123-QED}

\title{Vibronic dynamics in electron continuum - iterative solvers}

\author{Martina \'{C}osi\'{c}ov\'{a}}
 \altaffiliation[Current affiliation: ]{
Department of Applied Mathematics, \\ 
Faculty of Electrical Engineering and Computer Science, 
V\v{S}B~--~Technical University of Ostrava, Czech Republic}
\author{Jan Dvo\v{r}\'{a}k}%
 \altaffiliation[Current affiliation: ]{
 Chemical Sciences Division,
 Lawrence Berkeley National Laboratory,
 One Cyclotron Road,
 Berkeley, California 94720, USA}
\author{Martin \v{C}\'{\i}\v{z}ek}%
 \email{Martin.Cizek@mff.cuni.cz}
\affiliation{Charles University, Faculty of Mathematics and Physics, Institute of Theoretical Physics, \\
             V Hole\v{s}ovi\v{c}k\'{a}ch 2, 180 00 Prague, Czech Republic}


\date{\today}

\begin{abstract}
 We present a general two-dimensional model of conical intersection between 
 metastable states that are vibronically coupled not only directly but also indirectly through
 a virtual electron in the autodetachment continuum.
 This model is used as a test ground for design and comparison of iterative solvers for 
 resonance dynamics in low-energy electron-molecule collisions. 
 Two Krylov-subspace methods with various preconditioning schemes are compared. To 
 demonstrate the applicability of the proposed methods on even larger models, we also test 
 the performance of one of the methods  
 on a recent model of vibrational excitation of CO$_2$ by electron impact based on three 
 vibronically-coupled discrete states in continuum
 (Renner-Teller doublet of shape resonances coupled to sigma virtual state)
 including four vibrational degrees of freedom.
 Two-dimensional electron energy-loss spectra resulting from the electron-molecule 
 scattering within the models are briefly discussed.
\end{abstract}

\maketitle


\section{\label{sec:intro}Introduction} 

Despite of a long history of investigations \cite{BM1968,L1980,A1989}, 
the collisions of low-energy electrons with molecules still represent a fascinating and challenging 
field of study. By low energy we mean here the energy below the electronic-excitation threshold,
i.~e.\ the energy that does not exceed few units of electron volt. Even these low energies lead to 
many interesting phenomena like appearance of sharp structures in cross sections 
\cite{S1973,HRA2003}
or the possibility to select dissociation into different anionic fragments by tuning the energy 
\cite{PDM2005,IA2009}. 
This topic is both interesting for practical applications 
\cite{FEM2017}
and challenging for the theory even for small polyatomic molecules
\cite{rescigno_co2_2002,mccurdy_co2_2003,Formic2,Formic1,curik_2008,curik_diacetylene_2014,nag_nccn_2020}. 

In this paper we will focus on the process of the vibrational excitation in 
collision of an electron $e^-$ with a molecule $M$ initially in a vibrational state $|v_i\rangle$ 
\begin{equation}
 \label{eq:elmol}
 e^- + M(v_i) \to M^- \to e^- + M(v_f)
\end{equation}
mediated by one or several metastable anion states $M^-$. After the process, the molecule is left
in a final vibrational state $|v_f\rangle$. The total energy $E$ during the 
collision is conserved 
\begin{equation}\label{eq:Econserv}
  E = \epsilon_i + E_{v_i} = \epsilon_f + E_{v_f}, 
\end{equation}
where $\epsilon$ are electron energies and $E_v$ energies of vibrational states
of the molecule for the initial and the final states, before and after the collision.
This process is closely related to the process of photodetachment of an electron $e^-$
from a molecular anion $M^-$ 
\begin{equation}
 \label{eq:photod}
 \gamma + M^- \to (M^-)^* \to e^- + M(v_f),
\end{equation}
with initial energy $E$ of the system defined now by the energy of the photon $\gamma$
shone on the anion to excite it to the state $(M^-)^*$. The dynamics of both of these 
processes is driven by potential energy states of the negative molecular ion 
and their widths for decay into electronic continuum channels \cite{AMN2020}. 
The energies of the released electrons are sensitive to the relative position
of the anion and the neutral molecular states and the selection rules are different 
than for the radiative transitions \cite{gallup1986,gallup1993,CCA2015}.

The goal of this paper is to advance the detailed theory of the dynamics of the electron 
detachment from anions in such processes. In the development of the theoretical methods, we 
keep in mind the description of experiments that study in detail the energies of released 
electrons \cite{MCU2023,AMN2020,RNA2022}, and in particular, we calculate the two-dimensional 
electron energy-loss spectrum (2D EELS) for our model. The 2D electron loss spectroscopy
was pioneered by Currel and Commer 
\cite{RCC1988,CC1995} and further developed by Allan and collaborators \cite{RA2013}.
Up to now, a dozen of high-resolution spectra for different molecules have been measured 
\cite{RA2013,RA2015,ARG2016,ALP2018,RNZ2018,RKN2020,MSS2020,KKB2020,AMN2020}
but the detailed understanding of such spectra for polyatomic molecules is mostly lacking. 

In this paper, we present and test a general scheme for solving 
the nuclear dynamics of the negative ion formed in the collision of an electron 
with a polyatomic molecule. The scheme is tailored for a class of models 
that are inspired by the pseudo-Jahn-Teller model of Estrada, Cederbaum 
and Domcke \cite{ECD1986} with modifications meant to make it a more realistic 
model of real molecules. This approach combines a model of vibronic coupling 
of several anionic states expanded in low-order polynomials in vibrational 
coordinates close to equilibrium geometry of the neutral molecule with 
projection-operator approach to include the interaction of the anion 
discrete states with the electronic continuum. The model is rather flexible in 
adding states and vibrational degrees of freedom and 
the present scheme has been used to produce the results in our previous work 
on CO$_2$ \cite{the_letter,partI,partII}. In these papers we did not explain 
the methods and their performance in detail, a gap that is meant to be filled 
by this work.

We start Section II by reviewing the projection-operator approach to the dynamics 
of vibrational excitation in electron collisions with molecules. We then proceed by 
reminding the model of Estrada et al.~\cite{ECD1986} and propose its generalization
by including the vibronic coupling through the electron continuum in addition to 
the direct vibronic coupling present in the original model. 
This section is concluded by explaining the representation of the wave-function components
and the Hamiltonian in a basis constructed from neutral vibrational states. In Section III, we first 
introduce the used iteration methods and preconditioning schemes and then we discuss their
performance for the models. 
The section IV is devoted to a brief description of the obtained 2D spectra for the models
and we conclude by summarising the results in Section V. 

\section{\label{sec:theory}Theory}

 The vibrational and resonance dynamics in electron-molecule collisions has 
 been studied theoretically for a long time
(see, for example, one of the review papers \cite{L1980,GJ1986,D1991,Kniha_2012}). 
The direct brute-force approach is only tractable for small molecules \cite{SBF2021}
or for small deformations \cite{CC2003}. Number of approximate schemes have therefore
been developed: Born approximation, adiabatic-nuclei approximation, zero/effective range
or semi-classical approaches. In the present work we focus on the development 
of the numerical schemes for the projection-operator approach based on the 
existence of an intermediate anion state (or states) that is responsible 
for the coupling of the electronic and vibrational motion. The approach is 
often used in its approximate form --- the local complex potential approximation,
but it is known to fail in predicting interesting phenomena like Wigner cusps
or vibrationally excited Feshbach resonances. The nonlocal approach is well 
developed for diatomic molecules \cite{D1991,KB2020} but the attempts to use it for polyatomic 
molecules are scarce (see for example \cite{AF_2020}). In addition to bringing more 
degrees of freedom, the polyatomic molecules also exhibit interesting 
features like vibronic coupling of resonances, conical intersections and
exceptional points \cite{FSC2004,FC2004}.
Here, we follow the work of of Estrada, Cederbaum and Domcke \cite{ECD1986} (ECD86) 
and extend it to more general form of model functions and vibronic coupling. 
We start by presenting basic formulas resulting from the projection-operator formalism
(see \cite{D1991} for the comprehensive review of the approach). 
Then, we narrow the model to two vibrational degrees of freedom and two 
vibronically coupled discrete states.

\subsection{Nonlocal model for multiple discrete states in continuum}

The main idea of the nonlocal-discrete-state in continuum model is the assumption 
that the coupling of the electronic and vibrational degrees of freedom in 
the electron-molecule collision is mediated by one or a few discrete states
and after their removal from the electronic continuum using projection-operator
formalism of Feshbach \cite{feshbach_1962}, the electronic basis 
consisting of the discrete states and
the orthogonalized continuum is diabatic. The vibrational excitation or dissociative 
attachment then proceeds through capture into the discrete state. 

We 
define the projection operator 
\begin{equation}
    \mathcal{Q} = \sum_d \vert d\rangle\langle d\vert.
\end{equation}
as a sum over a set of discrete states $|d\rangle$ 
and the complementary operator
\begin{equation}
    \mathcal{P}=I - \mathcal{Q}
\end{equation}
projecting on the background continuum. 
The basis in the background part can be chosen as 
the states that solve the background scattering problem 
\begin{equation}\label{eq:bckgr}
  \mathcal{P}\mathcal{H}_{el}\mathcal{P}\vert\Phi_0,\epsilon\mu\rangle
  =(V_0+\epsilon)\vert\Phi_0,\epsilon\mu\rangle.
\end{equation}
Here, $V_0(\vec{q}\,)$ is the potential energy surface of the neutral molecule, 
i.\ e.\ the energy of the ground electronic state $|\Phi_0\rangle$
as function of the positions of the nuclei $\vec{q}$. 
Since we consider only low-energy electron scattering 
below threshold for the electronic excitation of the molecule, the state $\Phi_0$
is fixed and we will further omit it from the notation. The electron continuum states 
$|\epsilon\mu\rangle$ are thus uniquely described by the electron energy $\epsilon$
and some other quantum numbers collectively denoted by $\mu$
(typically angular momentum). All states $|d\rangle$
and $|\epsilon\mu\rangle$ thus form an orthogonal basis:
\begin{eqnarray}
  \langle d|d'\rangle                     & = & \delta_{dd'},\\
  \langle d|\epsilon\mu\rangle            & = & 0,\\
  \langle\epsilon\mu|\epsilon'\mu'\rangle & = & \delta_{\mu\mu'}\delta (\epsilon-\epsilon').
\end{eqnarray}

The electronic Hamiltonian in the $\mathcal{Q}$-space is described by a matrix
\begin{equation}\label{eq:H_d}
   \langle d|\mathcal{H}_{el}|d'\rangle
=
   V_0 \delta_{dd'} + U_{dd'}
\end{equation}
where all matrix elements depend on the molecular geometry, i.\ e.\ positions of nuclei $\vec{q}$.
The diagonal elements $V_d(\vec{q}\,)=V_0(\vec{q}\,)+U_{dd}(\vec{q}\,)$ represent the diabatic discrete-state
potentials and the off-diagonal part $U_{dd'}(\vec{q}\,)$ the direct vibronic coupling among
the states. The coupling between the discrete state $|d\rangle$ and the continuum $|\epsilon\mu\rangle$ is 
described by the coupling elements
\begin{equation}\label{eq:V_de}
   \langle d|\mathcal{H}_{el}|\epsilon\mu\rangle
=
   V_{d\epsilon}^{\mu}(\vec{q}\,).
\end{equation}
These elements represent the vibronic coupling
\footnote{
Note that we consider $V_{d\epsilon}^{\mu}$ to be real quantities. This 
is a reasonable assumption since the coupling can be made real for a single $\vec{q}$ by phase 
conventions and the dependence on $\vec{q}$ is supposed to be weak (diabaticity of the basis).}
of the discrete state to the continuum and they also 
lead to the second order vibronic coupling among the discrete states mediated by continuum 
as described below. 

This way, we parameterized the matrix elements given 
by Eqns.~\eqref{eq:bckgr}, \eqref{eq:H_d}, \eqref{eq:V_de} 
of the Hamiltonian $\mathcal{H}_{el}$ for the electron scattering from the molecule 
for each fixed position of the nuclei $\vec{q}$ by functions $V_0(\vec{q}\,)$, $U_{dd'}(\vec{q}\,)$ and
$V_{d\epsilon}^{\mu}(\vec{q}\,)$.
To describe the electron scattering 
from the molecule including the vibronic dynamics, we start from the definition of the vibrational 
states $|v\rangle$ of the target neutral molecule:
\begin{equation}\label{eq:H0}
  H_0|v\rangle = (T_N + V_0)|v\rangle = E_v|v\rangle,
\end{equation}
where $T_N$ is the kinetic-energy operator for the nuclei and $v$ is a set of quantum numbers that 
uniquely determine the vibrational states with energy $E_v$. It can be shown 
(see for example \cite{D1991}) 
that the vibronic motion of the anion is described by the effective Hamiltonian 
\begin{equation}\label{eq:H_ef}
    H_{ef} = H_0 + U + F,
\end{equation}
which is the matrix in the indices $d$, $d'$ and the operator in the space of vibrational degrees of freedom.
In the equation above, $H_0$ is the Hamiltonian operator for the vibrations of the molecule multiplied 
by unity matrix $\delta_{dd'}$ in the discrete state indices, $U$ is the matrix with the elements 
$U_{dd'}$ defined above and the operator $F$ describes the dynamical coupling of the discrete-state
space to the electronic continuum
\begin{equation}
    \label{eq:F_gen}
    F_{dd'}(E-H_0) = \\
        \sum_{\mu}\int_0^\infty 
           V_{d\epsilon}^{\mu}(\vec{q}\,)[E-H_0 - \epsilon + i\eta]^{-1} V_{d'\epsilon}^{\mu}(\vec{q}\,')
        \mathrm{d}\epsilon,
\end{equation}
where $\eta$ is a positive infinitesimal. This operator is a matrix 
in the discrete-state indices and nonlocal operator in the nuclear coordinate $\vec{q}$. 

The discrete-state contribution to the $T$-matrix for vibrational excitation by electron scattering 
in a continuum state $|\epsilon_i\mu_i\rangle$ from the initial vibrational state $v_i$ 
to final state $v_f$ and leaving in continuum state $|\epsilon_f\mu_f\rangle$ is given by
\cite{ECD1986}
\begin{equation}
    \label{eq:Td}
    T_{\mu_f v_f\leftarrow\mu_i v_i} =\sum_{dd'} 
        \langle v_f|V_{d\epsilon_f}^{\mu_f}
        [E-H_{ef}]_{dd'}^{-1}V_{d'\epsilon_i}^{\mu_i}|v_i\rangle
\end{equation}
and is closely related to the integral cross section for the vibrational excitation event
\begin{equation}
    \label{eq:crs}
    \sigma_{v_f\leftarrow v_i} = \frac{2\pi^3}{\epsilon_i}\sum_{\mu_i\mu_f}
    \left|T_{\mu_f v_f\leftarrow\mu_i v_i}\right|^2.
\end{equation}
Finally, to simulate the full 2D electron energy-loss spectra, we have to collect  
vibrational excitation cross sections for all accessible final states
\begin{equation}\label{eq:2Dsp}
    S(\epsilon_i,\Delta\epsilon) =
    \sum_{v_f} \sigma_{v_f\leftarrow v_i}(\epsilon_i)\rho(\Delta\epsilon-\Delta\epsilon_{v_f}),
\end{equation}
where $\rho(\epsilon)$ is the resolution function of the spectrometer (simulated here with 
a Gaussian function with full width at half maximum equaled to 10~meV, which is 
comparable to the values in the experiment \cite{RA2013}). The energy loss 
$\Delta\epsilon_{v_f}=E_{v_f}-E_{v_i}=\epsilon_i-\epsilon_f$ 
in each term in Eq.~\eqref{eq:2Dsp} is fixed by the energy conservation.
The function $S(\epsilon_i,\Delta\epsilon)$
gives the full experimental information in the electron energy-loss spectroscopy except 
for the angular resolution that can also be included \cite{partII} but it is not
of the interest in the present paper. 

\subsection{Pseudo Jahn-Teller model of Estrada et al.}

\begin{figure}[h!]
\centerline{\includegraphics[width=0.255\textwidth]{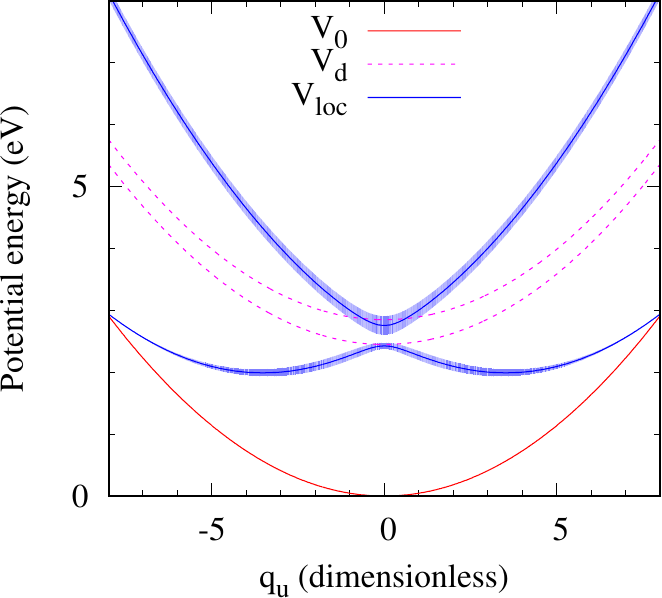}%
\hspace*{-3.5mm}\includegraphics[width=0.255\textwidth]{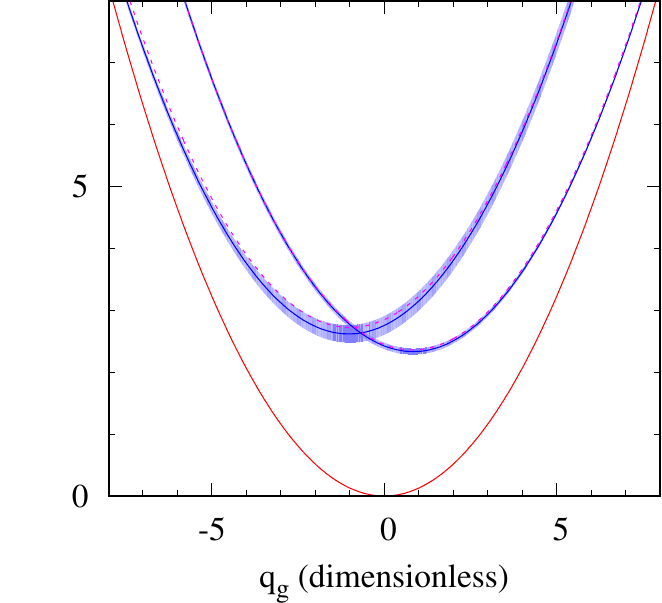}}
\caption{\label{fig:PotentECD}%
         Sections through the model potential energy surfaces in the $q_g=0$ (left) and $q_u=0$ (right)
         planes for the ECD86 model. Blue-shaded areas give the position and the width 
         of the fixed nuclei electronic resonance. }
\end{figure}
\begin{figure}
    \centering
    \includegraphics[width=0.47\textwidth]{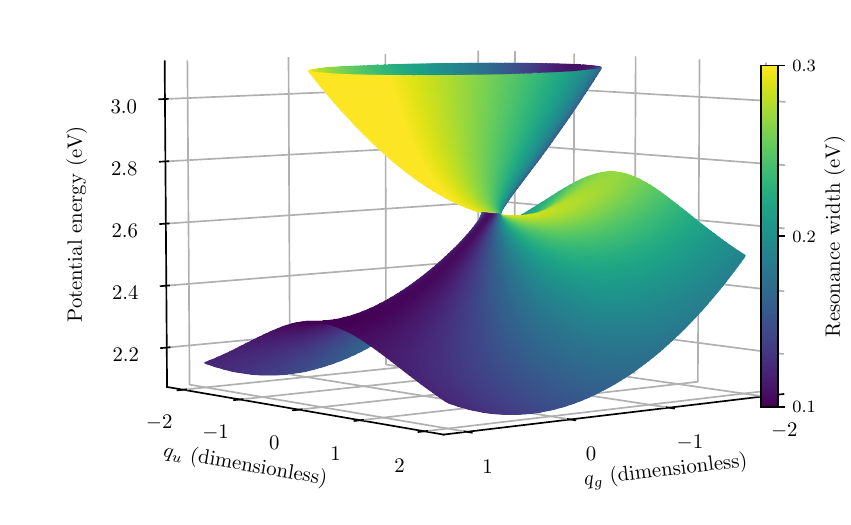}
    \caption{\label{fig:Pot3DECD86}%
             Perspective view of the potential energy manifold for the ECD86 model.
             The width (inverse lifetime) is marked by the color scale.
    }
\end{figure}
\begin{table}[b]
\caption{\label{tab:paramECD}%
The values of the parameters describing the pseudo Jahn-Teller 
model of Estrada et al.~\cite{ECD1986}.
}
\begin{ruledtabular}
\begin{tabular}{cdcd}
\textrm{Parameter}&
\textrm{Value\hspace*{-5mm}}&
\textrm{Parameter}&
\textrm{Value\hspace*{-5mm}}\\
\colrule
$\omega_g$    &  0.258 &  $\omega_u$ & 0.091 \\
$E_1$         &  2.45  &  $E_2$      & 2.85  \\
$\kappa_1$    & -0.212 &  $\kappa_2$ & 0.254 \\
$\lambda$     & 0.318  &             &       \\
$a_1$         & 0.086  &  $a_2$      & 0.186 \\
$b_1$         & 0.833  &  $b_2$      & 0.375 \\
$l_1$         & 2      &  $l_2$      & 1 
\end{tabular}
\end{ruledtabular}
\end{table}

Model by Estrada, Cederbaum and Domcke \cite{ECD1986} assumes a molecule with an Abelian group 
of symmetry. They consider two discrete states $d=1,2$ that transform according to different
irreducible representations of the symmetry group and are coupled vibronically through 
a nontotally symmetric vibrational mode $q_u$. They also consider excitation of another, 
totally symmetric mode $q_g$, so that the geometry of the molecule within the model is 
describe by a vector $\vec{q}=(q_g,q_u)$. 
The symmetry then dictates the structure of the matrix $U$:
\begin{equation}\label{eq:U}
    U =
    \left(\begin{array}{cc}
      E_1 + \kappa_1 q_g  &  \lambda q_u \\
        \lambda q_u       & E_2 +\kappa_2 q_g
    \end{array}\right).
\end{equation}
This is a completely general form of the dependence of the matrix $U$
on coordinates when the terms are restricted up to the first order in $\vec{q}$ and 
the symmetry requirements are taken into account. Similarly, we can expand the matrix of the 
discrete-state-continuum coupling. For simplicity, we consider only two partial waves
$|\epsilon\mu\rangle$ for $\mu=e,o$ representing one even and one odd linear 
combination of partial waves coupled to the discrete-state space. In principle, we 
could consider more partial waves but they could be decoupled from the problem 
by a unitary transformation, grouping thus partial waves into effective channels
with number of channels not exceeding the dimension of the $\mathcal{Q}$-space
\cite{M1968}. 
Estrada et al \cite{ECD1986} 
considered the coupling matrix $V_{d\epsilon}^{\mu}$ independent of the nuclear
coordinates (we are going to lift this restriction in the next section). The symmetry 
selection rules then forbid the coupling $V_{d\epsilon}^{\mu}$ to the different symmetry 
of the discrete state $d$ and the partial wave $\mu$. We therefore assume that the 
discrete state $d=1$ has even symmetry as the $\mu=e$ partial wave and $d=2$ has 
the symmetry of the odd partial wave $\mu=o$. Using Eq.~\eqref{eq:F_gen}, we then see 
that only the diagonal matrix elements $F_{11}(\epsilon)$ and $F_{22}(\epsilon)$
of the level-shift operator are nonzero. They can be generated from 
their imaginary parts (widths)
\begin{eqnarray*}
  \Gamma_{11}(\epsilon)&=&-2\,\mathrm{Im} F_{11}=2\pi|V_{1\epsilon}^{e}|^2,
\\
  \Gamma_{22}(\epsilon)&=&-2\,\mathrm{Im} F_{22}=2\pi|V_{2\epsilon}^{o}|^2
\end{eqnarray*}
by means of the integral transform $\mathcal{F}[\Gamma_{ii}(\epsilon)/2\pi]$ defined as
(compare Eq.~\eqref{eq:F_gen})
\begin{equation}\label{eq:IntF}
  \mathcal{F}[f(\epsilon)] = \int \frac{f(x)}{\epsilon - x + i\eta} {\rm d}x.
\end{equation}
This transform can be worked 
out analytically for the assumed form of the widths
\begin{equation}
    \Gamma_{dd}(\epsilon) = a_d \epsilon^{l_d+1/2}\exp (-b_d\epsilon)
\end{equation}
(see \cite{BEC1983}).
To complete the model description we must give the vibrational Hamiltonian $H_0$
of the neutral molecule. The model assumes simply harmonic 
vibrations 
\begin{equation}\label{eq:H0harm}
    H_0 = T_N + V_0 =
        -{\textstyle\frac{1}{2}}\omega_g\frac{\partial^2}{\partial q_g^2}
        -{\textstyle\frac{1}{2}}\omega_u\frac{\partial^2}{\partial q_u^2}
        +{\textstyle\frac{1}{2}}\omega_g q_g^2
        +{\textstyle\frac{1}{2}}\omega_u q_u^2.
\end{equation}

The vibrational eigenstates $|v\rangle$ satisfying Eq.~\eqref{eq:H0} with this harmonic Hamiltonian 
can be numbered by two quantum numbers $\nu = (n_g,n_u)$ and the vibrational energies 
are given by standard harmonic oscillator formula
$E_{\nu}=\omega_g (n_g + {\textstyle\frac{1}{2}})+\omega_u (n_u + {\textstyle\frac{1}{2}})$.

The numerical values of $\omega_i$ and the parameters defining direct coupling matrix $U$
and discrete-state-continuum matrix $V_e$ for the model studied in \cite{ECD1986} 
and used here for testing are given in Table~\ref{tab:paramECD}. Note that several variants
of the model were studied in \cite{ECD1986}. Here we study only the most complex form 
of the model with the values of parameters as given in the table. To visualize 
the character of the model, 
we show the one dimensional sections through the model potentials in 
Fig.~\ref{fig:PotentECD}. The functions shown are $V_0(\vec{q})$, 
$V_d(\vec{q})=V_0(\vec{q})+U_{dd}(\vec{q})$ 
and the local complex potential $V_{loc}(\vec{q})$. The last function is defined 
as the position $V_\mathrm{loc}=E_R-\frac{i}{2}\Gamma_R$ of the pole of the fixed 
nuclei S~matrix, which has to be located iteratively \cite{partI}.
The perspective view of the local complex potential 
$\operatorname{Re}V_\mathrm{loc}$ colored by values of $-2\operatorname{Im}V_\mathrm{loc}$ 
is also shown in Fig.~\ref{fig:Pot3DECD86}.

\subsection{\label{sec:gen_model}Generalized model with vibronic coupling with continuum states}

\begin{figure}[h!]
\centerline{\includegraphics[width=0.255\textwidth]{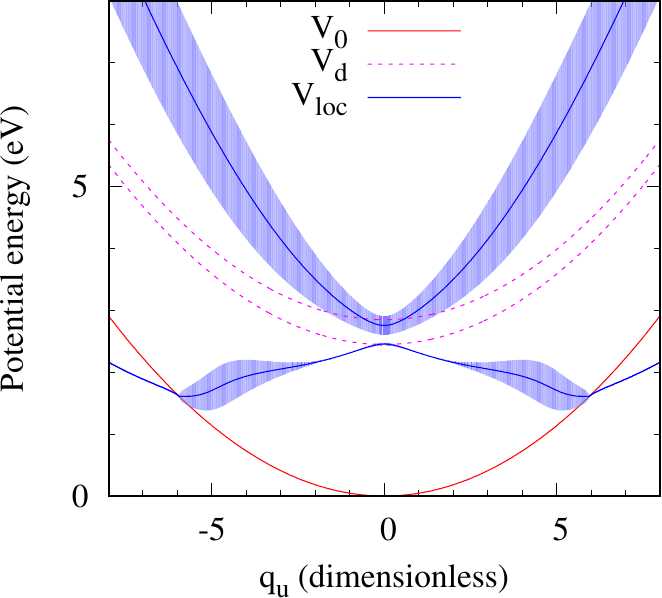}%
\hspace*{-3.5mm}\includegraphics[width=0.255\textwidth]{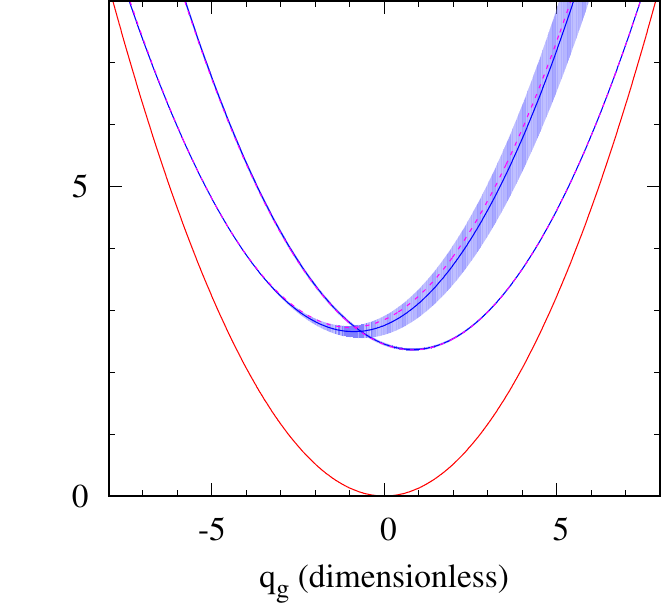}}
\caption{\label{fig:PotentB}
         Sections through the model potentials in the $q_g=0$ (left) and $q_u=0$ (right)
         planes for the new model.
}
\end{figure}
\begin{figure}
    \centering
    \includegraphics[width=0.47\textwidth]{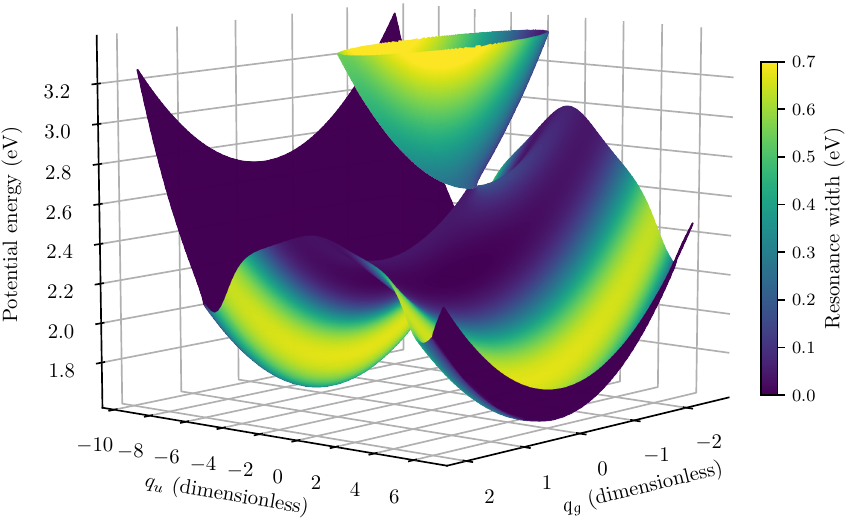}
    \caption{\label{fig:Pot3DNew}
             Perspective view of the potential energy manifold for the new model, colored 
             by the width of the resonance.
    }
\end{figure}
\begin{table}[b]
\caption{\label{tab:param.genECD}%
The values of the parameters describing the generalization of the model.
}
\begin{ruledtabular}
\begin{tabular}{cdcd}
\textrm{Parameter}&
\textrm{Value\hspace*{-5mm}}&
\textrm{Parameter}&
\textrm{Value\hspace*{-5mm}}\\
\colrule
$a_1^e$         & 0.07   &  $a_2^e$      & 0.1 \\
$b_1^e$         & 0.25   &  $b_2^e$      & 0.5 \\
$l_1^e$         & 0      &  $l_2^e$      & 0 \\
$a_1^o$         & 0.186  &  $a_2^o$      & 0.15 \\
$b_1^o$         & 0.375  &  $b_2^o$      & 0.8 \\
$l_1^o$         & 1      &  $l_2^o$      & 1 \\
$\lambda_1$     & 0.2    &  $\lambda_2$  & 0.1
\end{tabular}
\end{ruledtabular}
\end{table}

The vibronic model above assumes the most simple structure of the discrete-state-continuum 
coupling matrix $V_{\epsilon}=\{V_{d\epsilon}^{\mu}\}$ with row index $d$ and column index 
$\mu$:
$$
  V_{\epsilon}=
    \left(\begin{array}{cc}
        V_{1\epsilon}^e(\vec{q}\,) & V_{1\epsilon}^o(\vec{q}\,) \\
        V_{2\epsilon}^e(\vec{q}\,) & V_{2\epsilon}^o(\vec{q}\,)
    \end{array}\right)
=
    \left(\begin{array}{cc}
        \sqrt{\Gamma_{11}/2\pi} & 0 \\
                0             & \sqrt{\Gamma_{22}/2\pi}
    \end{array}\right).
$$
Here, we would like to go one step beyond the approximation of the coupling matrix by constant terms 
and we expand the matrix to the first order in the normal vibrational coordinates. 
This generalization is very useful in the description of the interaction of resonances through 
the electronic continuum that is switched off in the equilibrium geometry, but becomes nonzero
with deformation as, for example, in pyrrole molecule
\cite{RNR2022}
. This feature was also important 
ingredient of the model for CO$_2$ molecule \cite{the_letter, partI}.
We will further assume that the dependence of $V_{d\epsilon}(\vec{q}\,)$ on the electron energy $\epsilon$ 
and the normal coordinates is separable. Taking into account the symmetry of the system, we get
$$
  V_{\epsilon}=
    \left(\begin{array}{cc}
        f_{1}^e(\epsilon) [1+\lambda_1 q_g] & f_{1}^o(\epsilon) q_u \\
        f_{2}^e(\epsilon) q_u               & f_{2}^o(\epsilon) [1+\lambda_2 q_g]
    \end{array}\right),
$$
where the terms that couple a discrete state to the partial wave of different symmetry must be odd 
functions of $q_u$. We see that half of the total number of 12 terms (up to first order in $\vec{q}$) 
in the coupling matrix are zero due to the symmetry.
For the purposes of the testing of the numerical methods, we choose the same form 
of the energy dependence as in the original model:
\begin{equation}
    2 \pi |f_{d}^{\mu}|^2
    = a_d^{\mu} \epsilon^{l_d^{\mu}+1/2}\exp (-b_d^{\mu}\epsilon)
\end{equation}
with the values of the parameters given in Table \ref{tab:param.genECD}.

Using formula (\ref{eq:F_gen}), we see that the structure 
of the nonlocal level-shift operator $F(E-H_0)$ is much richer:
\begin{eqnarray} \nonumber
  F_{11} &=& (1+\lambda_1 q_g) \mathcal{F}[f_1^{e}f_1^{e}](1+\lambda_1 q_g) 
            + q_u \mathcal{F}[f_1^{o}f_1^{o}] q_u, 
\\ \nonumber
  F_{22} &=& q_u \mathcal{F}[f_2^{e}f_2^{e}] q_u  
            + (1+\lambda_2 q_g) \mathcal{F}[f_2^{o}f_2^{o}](1+\lambda_2 q_g), 
\\ \nonumber
  F_{12} &=& (1+\lambda_1 q_g) \mathcal{F}[f_1^e f_2^e ] q_u  
            + q_u \mathcal{F}[f_1^o f_2^o ](1+\lambda_2 q_g), 
\\ \nonumber
  F_{21} &=& q_u \mathcal{F}[f_1^e f_2^e ] (1+\lambda_1 q_g)  
            + (1+\lambda_2 q_g) \mathcal{F}[f_1^o f_2^o ] q_u,
\\ \mbox{~} \label{eq:FE}
\end{eqnarray}
where we used the integral transform (\ref{eq:IntF}) again.
The ordering of the terms that depend on $q_i$ with respect to $\mathcal{F}(\epsilon)$ 
must be kept because we substitute the operator $\epsilon = E-H_0$, which does not 
commute with the normal coordinates.

The potentials for the new generalized model are visualised in Figs.~\ref{fig:PotentB} 
and~\ref{fig:Pot3DNew}. Note that the structure of such conical intersections in 
continuum have been investigated in \cite{FC2004,FSC2004}. In accordance with their 
findings the potential manifolds shown in Figs.~\ref{fig:Pot3DECD86} and~\ref{fig:Pot3DNew}
do not intersect in a single point like regular conical intersections but in a line segment 
bounded by two exceptional points. The form of our model as given by Eq.~(\ref{eq:FE}) 
is more general than the expansion investigated in \cite{FC2004,FSC2004} because they 
studied a linear coordinate expansion of the width function $\Gamma$ whereas we prescribe 
the linear expansion of the coupling matrix $V_{\epsilon}$, which is more natural for 
the subsequent treatment of the dynamics. In our case, the linear form of the coupling matrix
produces also quadratic terms in widths in Eq.~(\ref{eq:FE}). When the quadratic terms are omitted,
we recover the form used in \cite{FC2004,FSC2004}. However, we can not 
omit these terms in the dynamics since it would distort the unitarity of the S~matrix.

\subsection{\label{sec:num_dynamics}Numerical representation of the dynamics}

For the numerical solution of the dynamics, we expand wave-function components 
in the harmonic oscillator basis $|\nu\rangle=|n_g,n_u\rangle$ associated with the model 
Hamiltonian of the neutral molecule (\ref{eq:H0harm}).
We first rewrite Eqs.~(\ref{eq:Td}), (\ref{eq:crs}) for the cross section as
\begin{equation}\label{eq:crs2}
  \sigma_{v_f\leftarrow v_i} = \frac{2\pi^3}{\epsilon_i}\sum_{\mu_i\mu_f}
    \left|
      \langle \Phi_{v_f}^{\mu_f}|\Psi_{v_i}^{\mu_i} \rangle
    \right|^2 ,
\end{equation}
where we defined auxiliary wave-functions $|\Phi_{v}^{\mu}\rangle$ with components
\begin{equation}
    \Phi^{v\mu}_{d,n_g,n_u} = \langle n_g,n_u|V_{d\epsilon}^{\mu}|v\rangle,
\end{equation}
and with energy according to the conservation law (\ref{eq:Econserv}), i.~e.\ %
$\epsilon = E - E_v$. The anion wave function $|\Psi_{v_i}^{\mu_i}\rangle$ satisfies
\begin{equation}\label{eq:Ax=b}
    (E-H_{ef}) |\Psi_{v_i}^{\mu_i}\rangle = |\Phi_{v_i}^{\mu_i}\rangle.
\end{equation}
In the harmonic oscillator basis, this equation represents a system of linear equations
for unknown components of the discrete-state wave function
\begin{equation}
    \Psi_{\alpha}\equiv
    \Psi_{d,n_g,n_u} = \langle d|\langle n_g,n_u|\Psi_{v_i}^{\mu_i}\rangle,
\end{equation}
where we introduced a compound index $\alpha\equiv(d,n_g,n_u)$.
Using this notation the matrix of this system $A_{\alpha,\alpha'}$ reads
\begin{equation}\label{eq:Amatrix}
    A_{\alpha,\alpha'} = \langle n_g,n_u| (E-H_{ef})_{dd'}|n_g',n_u'\rangle
\end{equation}
and the scalar product in Eq.~(\ref{eq:crs2}) can be written as the sum 
$\sum_{\alpha}\equiv\sum_d\sum_{n_g}\sum_{n_u}$ over the components of 
$\Phi^{v_f\mu_f}_{\alpha}$ and $\Psi_{\alpha}$. 
We cut of the basis in each dimension keeping the states
$|n_g\rangle$ for $n_g=0,1,...,N_g-1$ and 
$|n_u\rangle$ for $n_u=0,1,...,N_u-1$. The states $\Psi_{\alpha}$ are thus represented 
by $N=2N_g N_u$ component vectors and $A$ is a $N\times N$ matrix.

\begin{figure*}[th]
\includegraphics[width=1.07\textwidth]{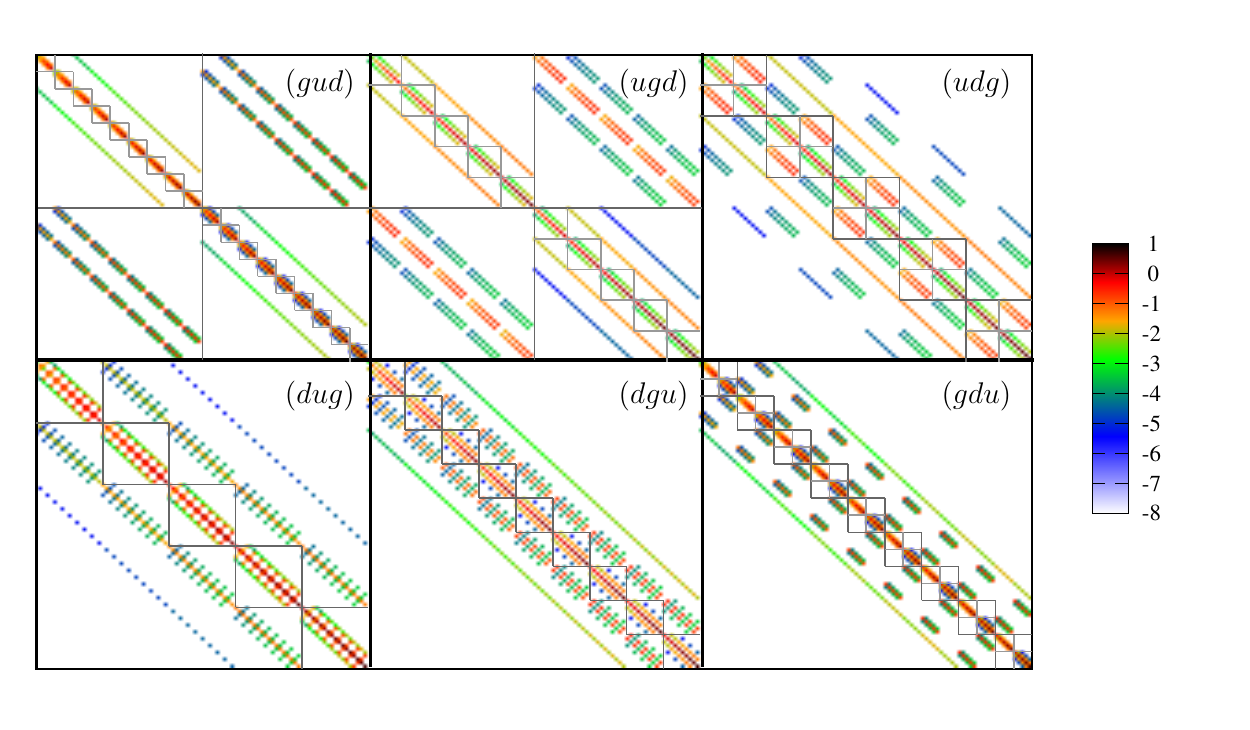}
\caption{\label{fig:Astruc} Structure of the matrix $A$ of the generalized model depending on the ordering of the basis,
         where the order is given by three letters $d$, $g$, and $u$ representing the discrete states and 
         vibratonal modes $q_g$ and $q_u$. The first index changes the fastest.
         The size of the matrix elements is represented by color in a logarithmic scale.
         The block-diagonal parts used for preconditioning are shown by the squares and for some orderings, the
         blocks corresponding to the discrete state are also indicated.}
\end{figure*}

For the solution of Eq.~(\ref{eq:Ax=b}) in the original model, Estrada et al.~\cite{ECD1986} 
devised a specially tailored method based 
on the block-tridiagonal structure of the matrix $A$. Our aim in this paper is to develop a more 
general method 
capable of solving a larger class of models
and test it both on the original model and on our generalization. 
The matrix $A$ is large but sparse. From the character of the problem, it is also complex symmetric, but not Hermitian. The structure of the matrix $A$ depends on the order 
of the basis vectors as illustrated in Fig.~\ref{fig:Astruc} for the generalized model.

\section{Krylov-subspace iteration methods\label{sec:Krylov}}

The Krylov-subspace iteration methods are well suited for solving Eq.~(\ref{eq:Ax=b}).
The main idea of all of the Krylov-subspace methods is that they solve a linear system  
$$Ax=b$$
iteratively producing a sequence $x_0,x_1,x_2,\ldots,x_n$ of approximations of the solution vector~$x$ in 
the Krylov-subspace, i.~e.\ in the space
$$\mathcal{K}_n\left(A ; r_0 \right) = span\left\{r_0,A r_0,A^2 r_0,...,A^{n-1} r_0\right\}.$$
where $r_0 = b-A x_0$ denotes the initial residual vector.
The methods differ by the definition of the "ideal" approximation of the solution $x_n$ within 
the Krylov subspace and usually proceed by application of simple recursive formulas. 
To produce the Krylov subspace we only need to implement the matrix multiplication $Aw$ for an
arbitrary vector $w$. The choice of the harmonic basis is very convenient for implementation 
of this matrix multiplication. From Eqs.~(\ref{eq:H_ef}), (\ref{eq:U}), (\ref{eq:FE}), we can see  
that the multiplication by the matrix $A$ can be decomposed to successive multiplications 
by energy-dependent diagonal factors, for example,
\begin{eqnarray*} 
  [(E-H_0)w]_{\alpha}               & = & (E-E_{n_g,n_u})w_{d,n_g,n_u},\\
  {[\mathcal{F}(E-H_0)w]}_{\alpha}  & = & \mathcal{F}(E-E_{n_g,n_u})w_{d,n_g,n_u}.
\end{eqnarray*}
and by operators of the coordinates $q_g$ and $q_u$
\begin{eqnarray*}
  [q_g w]_{\alpha} &=& {\textstyle\frac{1}{\sqrt{2}}}
  \left(w_{d,n_g{-}1,n_u}\sqrt{n_g}+w_{d,n_g{+}1,n_u}\sqrt{n_g+1}\right)\\ 
  {[q_u w]}_{\alpha} &=& {\textstyle\frac{1}{\sqrt{2}}}
  \left(w_{d,n_g,n_u{-}1}\sqrt{n_u}+w_{d,n_g,n_u{+}1}\sqrt{n_u+1}\right).
\end{eqnarray*}
These operations can be implemented very efficiently. Note that all energy factors and square roots 
can be precalculated and stored before starting the iteration process. Compared to a matrix-vector 
multiplication, which requires $O(N^2)$ operations for full matrices, the above procedure 
requires only $O(N)$ operations (with the operation count being approximately three times larger 
for the generalized model due to more complicated structure of the operator $F$).
The efficiency of the method of solution 
of Eq.~(\ref{eq:Ax=b}) is then given by the rate of convergence of the sequence 
$x_n$ to the solution, which is judged by monitoring the size of the norm of 
the residuum $\left\Vert r_n\right\Vert=\left\Vert b-Ax_n\right\Vert$. 
In the following tests we stop the iterations when the value 
$\left\Vert r_n\right\Vert<10^{-6} \left\Vert b \right\Vert$ 
is reached.

\subsection{Methods of interest}

Saad and Schultz \cite{Saad-Schultz} developed the generalized minimal residual (GMRES) method,
one of the most widely used Krylov-subspace methods.
This method constructs an orthonormal basis $v_1, v_2, \ldots , v_{n}$ of the $n$-th Krylov subspace $\mathcal{K}_n\left(A ; r_0 \right)$ using the Arnoldi algorithm that can be written in the matrix form as
$$A V_n = V_{n+1}\tilde{H}_n, \quad V_n=[v_1|v_2|\cdots|v_n]$$,
where $\tilde{H}_n \in \mathbb{C}^{(n+1)\times n}$ is an upper Hessenberg matrix.
The approximation $x_n$ of the solution in each step is given by the condition that
the residual vector $r_n = b-A x_n$ satisfies the optimality property:
\begin{equation}\label{eq:GMRES_min}
\left\Vert r_n \right\Vert = \min\limits_{x_n\in x_0+\mathcal{K}_n\left(A ; r_0 \right)}\left\Vert b-A x_n \right\Vert.
\end{equation}
If we write $x_n = x_0 + V_n y_n$, this condition leads to the $(n+1)\times n$ least-square problem for $y_n$:
$$y_n = \argmin\limits_{y\in\mathbb{C}^{n}}\left\Vert \left\Vert r_0\right\Vert e_1 - \tilde{H}_n y \right\Vert,$$
which has to be solved in every iteration.  

An advantage of the GMRES method is that it can be used for any matrix $A$ with no other special properties required 
than the regularity. On the other hand, each new basis vector at every iteration step has to be orthogonalized to 
all previous vectors. Thus, the size of the matrix $V_n$ grows during the iteration process 
and if the method does not converge quickly, the storage space and the time needed for each 
step grows significantly.
Note that the matrix A of our system of equations is complex symmetric (i.\ e.\ not Hermitian). 
For this reason, the Conjugate gradient method (which, unlike the GMRES method, uses short-term 
recurrences to construct the basis of the Krylov subspace, making it much less computationally demanding) 
can not be used to solve it. 
It is known \cite{Faber_Manteuffel1984} that for non-normal
matrices it is not possible to define an `optimal' iterative process (i.~e.\ a process that minimizes the residual or certain norm of the error over the Krylov subspace) that constructs the basis of Krylov subspace using short-term recurrences \cite{Liesen_Strakos_2013}. 
For complex symmetric matrices
van der Vorst and Mellisen~\cite{Vorst} presented an alternative way to define an iterative process based on three-term recurrences and derived the conjugate orthogonal conjugate gradient (COCG) method. Setting $v_0 = r_0 = b-Ax_0$, the basis of the Krylov subspace is constructed using the recursive formula
$$v_{n+1} = Av_n + \alpha_n v_n + \beta_n v_{n-1},$$
where $\alpha_n$ and $\beta_n$ follow from the conditions 
$\left\langle v_{n+1}|v_n\right\rangle_S = 0$ 
and 
$\left\langle {v}_{n+1}|v_{n-1}\right\rangle_S = 0$.
These conditions are analogous to those that define the Conjugate gradient method, in which, however, we have replaced the standard scalar product
with the symmetrized bilinear form 
$$\left\langle a|b\right\rangle_S = \left\langle a^*|b\right\rangle, \quad a, b \in \mathbb{C}^N.$$
Note that the complex conjugation in the left argument cancels the complex conjugation in
the standard definition of the scalar product. The bilinear form 
$\left\langle \cdot |\cdot \right\rangle_S$ is not therefore positive definite but it preserves 
the symmetry of the matrix $A$. 
This process ensures that vectors $v_0, v_1, \ldots, v_n$ satisfy the conjugate orthogonality property
(vectors $a, b\in \mathbb{C}^N$ 
are conjugate orthogonal if 
$\left\langle {a}|b\right\rangle_S = 0$). 
The whole iterative process is thus analogous to the conjugate gradient method 
but the convergence after $N$ steps ($N$ being the dimension of the matrix $A$) is not guaranteed in exact 
arithmetic.
In addition, the iterations do not have to converge at all since the symmetrized product 
can be zero (or very small) even for nonzero vectors.
In practice, however, the convergence is usually achieved. Moreover, (especially with proper preconditioning) it is often rather fast.

\subsection{Preconditioning}

In this paper, by preconditioning we understand the transformation of the 
original linear system $Ax=b$ into an equivalent problem (see, for example, \cite{Saad_2003})
\begin{equation}
  M^{-1}AM^{-T} y = M^{-1}b, \qquad \mbox{for~} y=M^{T}x,
\end{equation}
with a regular matrix $M$. This way, the preconditioning preserves the symmetry of the matrix $A$. 
As a rule of thumb, a good preconditioner~$M$ is represented by some fast invertible 
approximation of the original matrix $A$, but there are no exact guidelines 
for choosing the ideal matrix $M$ ensuring a fast convergence for the transformed problem.
There is a large variety of preconditioners
known in the literature but usually they are proposed for specific problems with matrices 
$A$ having some special properties. We tested some of the preconditiononing options 
but we found that the most simple methods work best for the present problem
\cite{Bcl_Martina,Ms_Martina}. In the following, 
we discuss different possibilities of block-diagonal preconditioning.
For this choice we define the matrix $M$ as a block diagonal section of the original 
matrix $A$. 
Depending on the structure of the blocks, the individual blocks
can either be inverted directly to construct $M^{-1}$ or a banded structure of the blocks 
can be used. The exact structure of the block depends on the ordering of the 
basis (see Fig.~\ref{fig:Astruc}). We thus define three different preconditioners 
$M^{dg}$, $M^{gu}$, $M^{du}$ with small diagonal blocks of the sizes $N_u\times N_u$,
$2\times 2$ and $N_g\times N_g$, respectively, and three preconditioners 
$M^{d}$, $M^{g}$, $M^{u}$ with large blocks of sizes 
$N_gN_u\times N_gN_u$, $2N_u\times 2N_u$ and $2N_g\times 2N_g$.
To be more specific, in the $(dug)$ ordering of the basis functions, the small-block preconditioning
matrix $M^{ug}$ has $N_uN_g$ blocks $M^{(n_u,n_g)}$ of the size $2\times 2$ 
with the matrix elements 
\begin{equation}
  M^{(n_u,n_g)}_{d,d'}=A_{(d,n_g,n_u),(d',n_g,n_u)}
\end{equation}
and the large-block preconditioning matrix $M^{g}$ has $N_g$ blocks $M^{(n_g)}$ of the size 
$2N_u\times 2N_u$ with the matrix elements 
\begin{equation}
  M^{(n_g)}_{d n_u,d' n_u'}=A_{(d,n_g,n_u),(d',n_g,n_u')}.
\end{equation}

To apply the preconditioning we need to act with $M^{-1}$ on a vector $w$
with components $w_{d,n_g,n_u}$. This can be done block by block;
for example, to apply the preconditioner $M^{u}$ we invert each 
block $M^{(n_u)}$ using the $LL^T$ (or $LDL^T$) decomposition and act on the section of 
$w_{d,n_g,n_u}$-vector for this fixed $n_u$
\begin{equation}
  [{M^{(n_u)}}^{-1} w]_{d,n_g,n_u} = 
    \sum_{d',n_g'} [{M^{(n_u)}}^{-1}]_{dn_g,d'n_g'}w_{d',n_g',n_u}.
\end{equation}
This is repeated for each $n_u$. Note that the inverting of the preconditioning 
matrices $M^{(n_u)}$ should be done just once and stored before starting iterations.
Moreover, in practice the $L$ (and $D$) matrix is 
stored in memory instead of ${M^{(n_u)}}^{-1}$.

\begin{figure}[h!]
\includegraphics[width=0.46\textwidth]{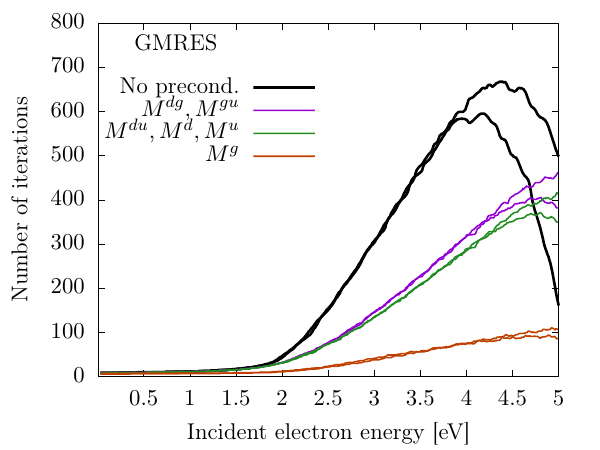}
\includegraphics[width=0.46\textwidth]{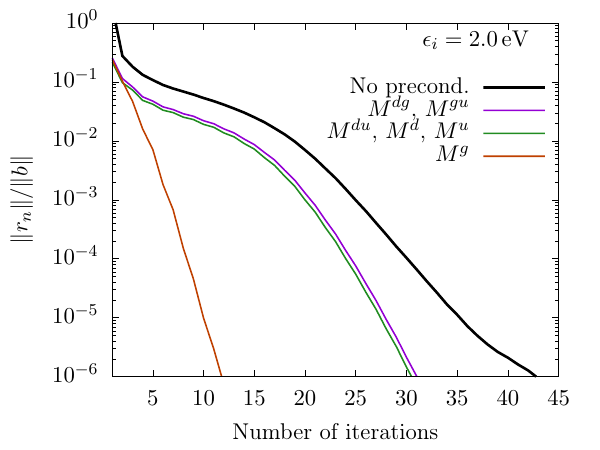}
\includegraphics[width=0.46\textwidth]{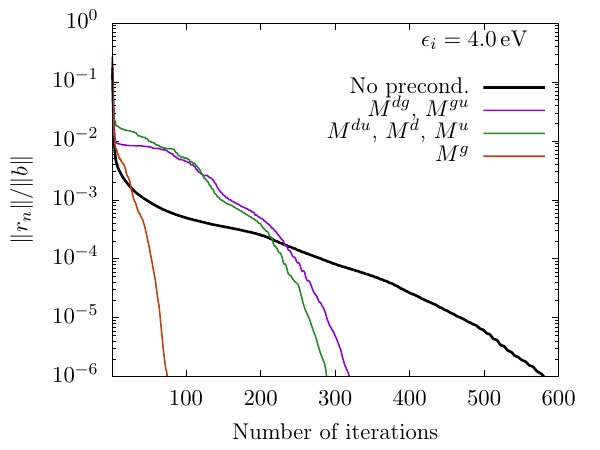}
\caption{\label{fig:ECD_GMRES}
          Convergence of GMRES method for the model of Estrada et al.\ \cite{ECD1986} for
          various preconditioners.
          Number of iterations needed for each electron energy (top) and 
          convergence of residuum for $\epsilon_i=$2~eV and 4~eV (bottom two panels).
          The top part shows results for both right-hand sides $\mu=e,~o$ of the linear system;
          bottom two parts display just $\mu=e$.}
\end{figure}

\begin{figure}[th]
\includegraphics[width=0.46\textwidth]{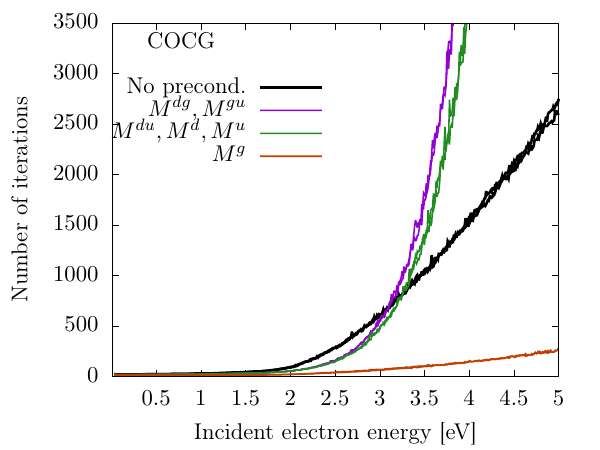}
\includegraphics[width=0.46\textwidth]{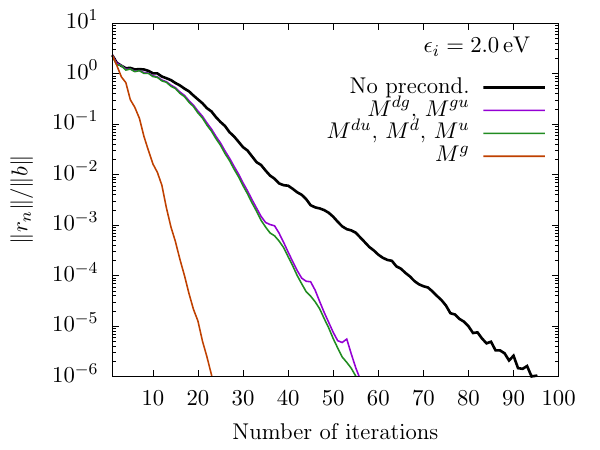}
\includegraphics[width=0.46\textwidth]{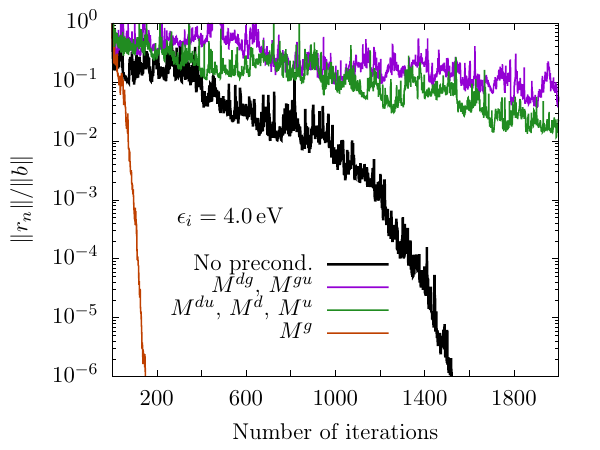}
\caption{\label{fig:ECD_COCG}
          Convergence of COCG method for the model of Estrada et al.\ \cite{ECD1986},
          see also caption of Fig.~\ref{fig:ECD_GMRES} for details.}
\end{figure}

\subsection{Numerical testing}

We applied the above described methods GMRES and COCG to solve the system (\ref{eq:Ax=b})
with the matrix (\ref{eq:Amatrix}) for the original ECD86 model and for our generalization
of the model. 
The performance of each method for different preconditioning is discussed separately 
for the two models in the next two paragraphs. The last paragraph also discusses 
the performance of the COCG method for a realistic model that describes inelastic 
electron scattering from the CO$_2$ molecule.

\paragraph{Perfomance of the methods for ECD86 model}
is shown in Fig.~\ref{fig:ECD_GMRES} and Fig.~\ref{fig:ECD_COCG}.
Each of the figures is devoted to one of the methods comparing different preconditioning 
schemes. The top graph summarises the number of iterations needed for convergence for all energies 
and bottom two graphs demonstrate the decrease of the residuum norm for two selected energies 
$\epsilon_i=$2eV and 4eV. 
The different preconditioning methods are shown with different colors. The curves of the same color 
correspond to the two different right-hand sides $\mu_i=o,e$ in Eq.~(\ref{eq:Ax=b}).

Let us first focus on graphs at the top of Fig.~\ref{fig:ECD_GMRES} showing the performance 
of the GMRES method. The method converges rather well (less then 700 iterations) even without any
preconditioning. The convergence is extremely fast below 2~eV (several dozens of iterations) 
but gets slower above this energy with maximum around 4eV. This is related to the spectrum 
of the anion. The electron with energy below 2~eV 
does not have enough energy to populate vibrational states of the anionic potential.
The process of the electron scattering is therefore 
almost elastic, which means that the wave function is not much 
perturbed with respect to the initial state used for starting the iterations. 
Above this energy, the dynamics is much richer, which is reflected in the increased 
number of iterations needed to reach the converged wave function. For the most of the energies
in the range of interest, the preconditioning reduces the number of iterations considerably.
The least efficient preconditioning matrices $M^{dg}$, $M^{gu}$ 
(overlapping curves in Fig.~\ref{fig:ECD_GMRES}) include only diagonal portion of the matrix $A$ 
and are therefore numerically very cheap to implement. 
The preconditioner $M^{du}$ includes also terms proportional to 
coupling constants $\kappa_1$, $\kappa_2$. For the ECD86 model, there are no terms
in the matrix $A$ added by increasing the size of preconditioner to $M^{d}$ and $M^{u}$, 
the convergence curves thus overlap for these three preconditioners. The 
best results are obtained with the preconditioning matrix $M^{g}$ which has blocks 
of size $2N_u\times 2N_u$ and includes terms proportional to $\lambda$ in 
Eq.~(\ref{eq:U}). The convergence of residuum norm in the lower part 
of Fig.~\ref{fig:ECD_GMRES} shows a difference in behavior of different preconditioned 
methods. While the best method with $M^{g}$ preconditioner converges exponentionally 
for all energies, there is kind of plateau in the other methods and the iterations
without preconditioning even became more efficient at high energies. 

The behavior of the COCG method (Fig.~\ref{fig:ECD_COCG}( is different in several aspects.
The overall number of iterations is approximately three time larger (for unpreconditioned
iterations) but we have to keep in mind that the COCG method is much simpler with 
computational demands constant over the course of the iterations. For GMRES,
the computational demands for one iteration grows quadraticaly 
with the number of iterations.
The COCG method does not have the minimization property 
(\ref{eq:GMRES_min}). This is reflected in the shape of the convergence curves
(two bottom graphs in Fig.~\ref{fig:ECD_COCG}). Unlike in similar curves for GMRES, 
here the residuum can locally grow although in general it finally converges 
to zero. The efficiency of the different preconditioning schemes is similar like 
in GMRES, although for higher energies only the $M^g$ preconditioner is useful. 

\paragraph{Performance of the methods for generalized model.}
\begin{figure}[th]
\includegraphics[width=0.46\textwidth]{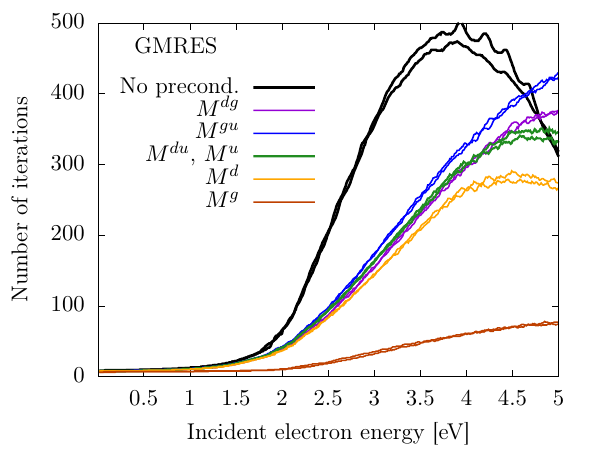}
\includegraphics[width=0.46\textwidth]{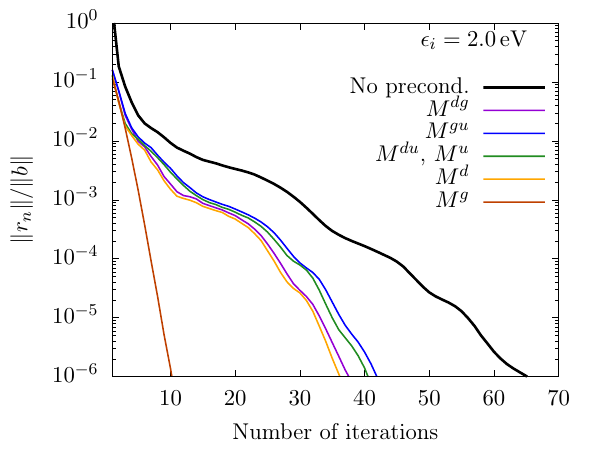}
\includegraphics[width=0.46\textwidth]{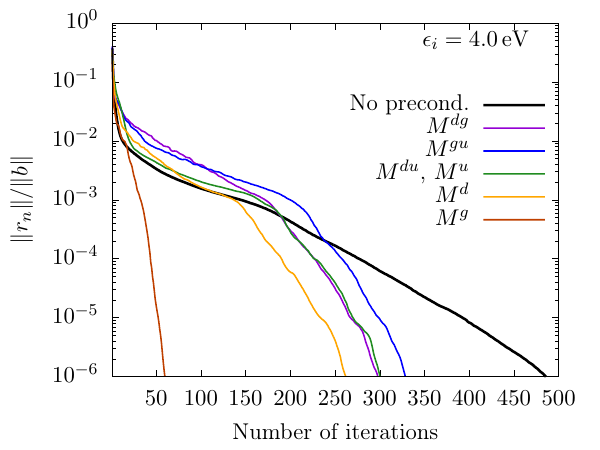}
\caption{\label{fig:NewMod_GMRES}
          Convergence of GMRES method for the generalized model,
          see also caption of Fig.~\ref{fig:ECD_GMRES} for details.}
\end{figure}
\begin{figure}[th]
\includegraphics[width=0.46\textwidth]{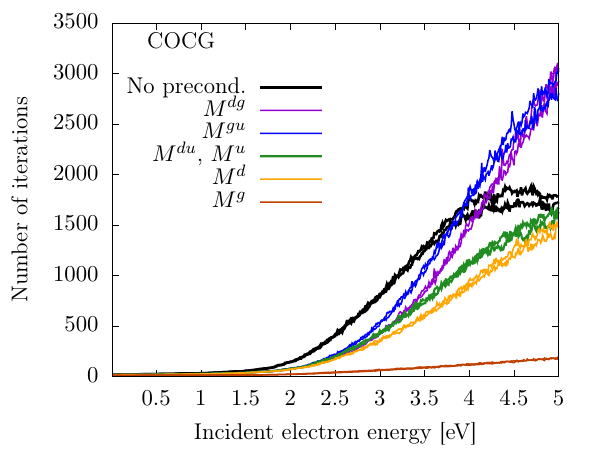}
\includegraphics[width=0.46\textwidth]{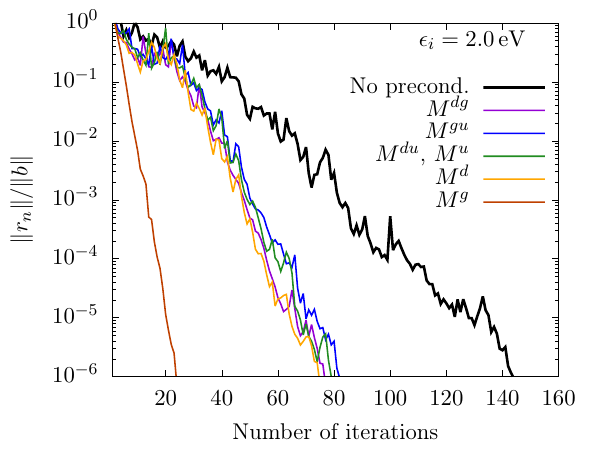}
\includegraphics[width=0.46\textwidth]{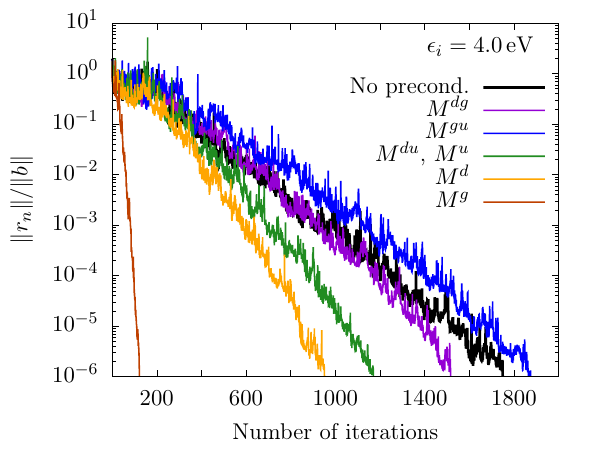}
\caption{\label{fig:NewMod_COCG}
          Convergence of COCG method for the generalized model,
          see also caption of Fig.~\ref{fig:ECD_GMRES} for details.}
\end{figure}

The generalized model has a more complicated structure (\ref{eq:FE}) of the level-shift 
operator $F(E)$, which is reflected in a more complicated structure of the matrix $A$, 
see Fig.~\ref{fig:Astruc}.
Surprisingly, the iteration methods converge faster with this matrix. There are no 
clear criteria relating rigorously the structure of the matrix to the speed of convergence. 
We believe that the faster convergence here may be related 
to the fact that operator $F$ in generalized model increases diagonal elements
of the matrix $A$. Apart from a little bit faster convergence, the graphs in 
Fig.~\ref{fig:NewMod_GMRES} for the GMRES method in the new model look qualitatively 
similar as for ECD86 model. The norm of the residuum is monotonously decreasing for 
all methods and the preconditioner $M^g$ is again the most efficient. The individual 
preconditioners now lead to different convergence rates, because all choices of the diagonal
blocks are distinct for the richer structure of $A$. The exception 
is the equivalence of $M^{du}$ and $M^u$ preconditioning (the green and yellow
lines are overlapping in the graphs). This can be nicely understood from the structure
of the matrix $A$ depicted in Fig.~\ref{fig:Astruc}. We see that the large 
and small black diagonal boxes in the bottom right matrix differ by a blank area
of zero matrix elements. 

The faster convergence for the new model is even more apparent for the COCG method
in Fig.~\ref{fig:NewMod_COCG}. Now all preconditioning schemes except for 
$M^{dg}$ and $M^{gu}$ are faster than direct iterations. 

To conclude the numerical experiment section, we would like to add a few 
notes on the implementation. Even without utilizing the structure 
of the matrix $A$, we have got by one order of magnitude faster calculation 
of the spectra utilizing the Krylov-subspace iteration methods as compared to 
a direct solver. Optimizing the matrix-vector multiplication using 
the structure of the matrix $A$ explained at the beginning of Sec.~\ref{sec:Krylov}
leads to the another order of magnitude speed up. From the previous 
examples, we see that the proper choice of preconditioning leads 
to the decrease of number of iterations needed for convergence by another 
one order of magnitude for both models and both methods. 

\paragraph{Performance for model of $e^-$+CO$_2$.}

In the final part of this section, we discuss our earlier work~\cite{the_letter, partI, partII} 
on the electron collisions with the carbon dioxide (CO$_2$) molecule 
in the context of the present paper. The vibronic coupling model for the $e+\mathrm{CO}_2$ 
system~\cite{partI} follows the general approach presented here in Sec.~\ref{sec:gen_model},
however, the model is more complex. We considered the nuclear motion 
within the full four-dimensional vibrational space in combination with three electronic
states ($^2\Sigma_g^+$ virtual state and two components of $^2\Pi_u$ shape resonance),
which are coupled upon bending of the molecule. The Hamiltonian 
is thus a $3\times 3$ matrix in the electronic space and we did not restrict its
elements only to the first order in the normal coordinates (some of the elements were
expanded up to the fourth order). Additionally, the three discrete states were coupled to
four electron partial waves. The vibrational dynamics is described analogically
to the scheme given in Sec.~\ref{sec:num_dynamics} but there are four vibrational
indices instead of two. The vibrational basis was constructed from products of eigenfunctions 
of 1D harmonic oscillators for symmetric and stretching modes and eigenfunctions of 
2D harmonic oscillator expressed in polar coordinates for the two-dimensional bending mode.

\begin{figure}
    \centering
    \includegraphics[scale=1.0]{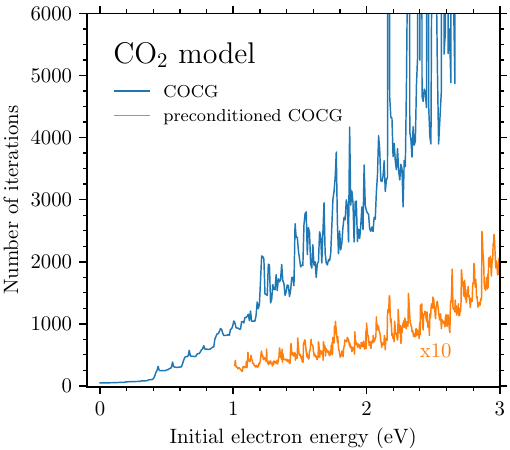}
    \caption{\label{fig:co2_convergence} 
    Number of iterations needed to solve the Schr\"odinger equation for the
    $e+\mathrm{CO}_2$ system using the COCG method without and with preconditioning. In the latter case,
    the curve is multiplied by a factor of 10.
    }
\end{figure}

Using the COCG method without any preconditioning, the number of iterations needed to reach the 
convergence with the stopping criterion of $10^{-3}$ (sufficient to obtain converged cross sections)
rapidly grows with the electron energy,
see Fig.~\ref{fig:co2_convergence}. For energies above 3~eV, even $2\times 10^5$ iterations were insufficient
to reach the convergence, therefore, a suitable preconditioning is essential.

The slow rate of convergence or no convergence at all is caused by the coupling of the discrete states 
through the bending mode. The stretching modes do not affect the convergence much since we found that the COCG
method converges badly even for the case where we did not consider the stretching modes.\footnote{We can easily freeze
a vibrational mode by considering only the ground state as the basis within this mode and setting
all relevant model parameters to zero.} 
Thus, taking a block-diagonal preconditioner where blocks contain the discrete states and the two-dimensional
bending was a natural choice. Such a preconditioner is analogous to the preconditioner $M^g$ that performs the best
for the ECD86 model and its generalization. In the case of CO$_2$, around 200 iterations were sufficient to reach the
convergence for initial electron energy of 3~eV, see Fig.~\ref{fig:co2_convergence}.

\begin{figure}[th]
\includegraphics[width=0.49\textwidth]{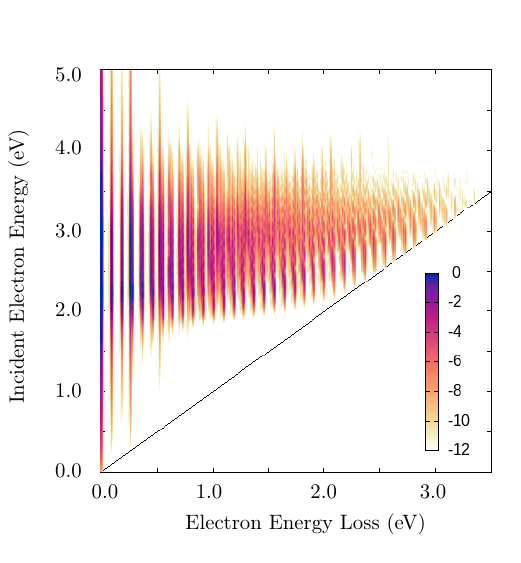}
\includegraphics[width=0.49\textwidth]{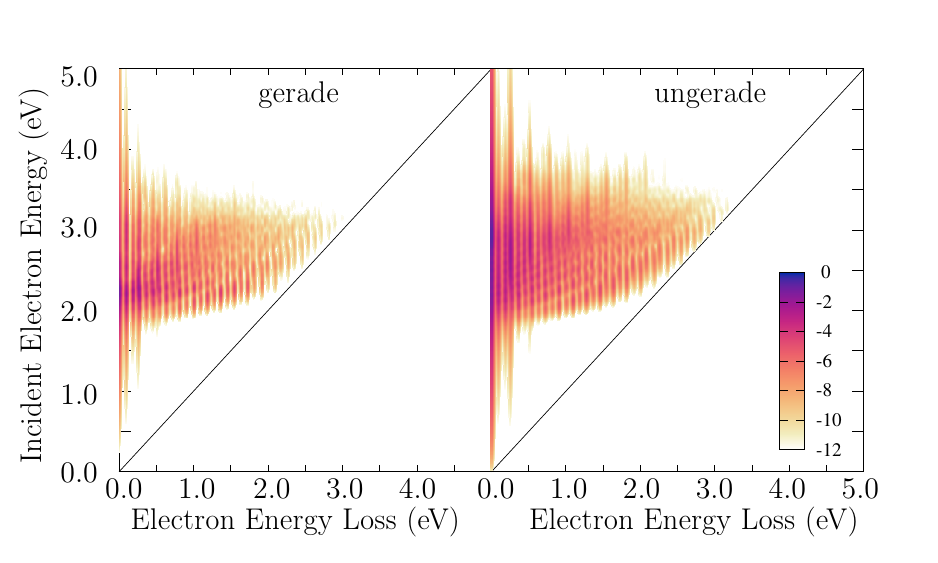}
\caption{\label{fig:EELS_A} 
          2D electron energy-loss spectrum for ECD86 model (top) and its 
          decomposition to the gerade and ungerade symmetries (bottom). The intensity 
          of spectrum is shown in a logarithmic scale.}
\end{figure}

\section{Discussion of resulting spectra for test models}

It is not purpose of this paper to study in detail the calculated spectra and their interpretation. 
This will require a detailed analysis of the final-state distribution and shape of the individual 
components of the wave function in the coordinate representation and its relation to the shape 
of potentials and also study of the dependence of the results on the model parameters. It is 
quite voluminous work that deserves a separate paper. We would also like to identify specific 
molecules that can be treated with the model of the current setup or a proper generalization. 
We already published the generalization of the model~\cite{partI} needed to describe the resulting 
spectra for the CO$_2$ molecule~\cite{the_letter} and performed the detailed analysis~\cite{partII}
including the final-state distribution, the wave functions and decomposition of spectra due 
to contribution of components of different symmetry.

In the following, we just show and briefly describe the 2D spectra for the ECD86 model 
(which were not subject of their original paper) and for our new generalization of the
model. We also separate the contribution of the two right hand sides in 
Eq.~(\ref{eq:Ax=b}) corresponding to the gerade and ungerade symmetry.

\subsection{2D spectrum for ECD86 model}

The calculated 2D spectrum for the ECD86 model is shown in Fig.~\ref{fig:EELS_A}.
The intensity given by Eq.~\eqref{eq:2Dsp} is plotted as a function of both energy loss 
$\Delta\epsilon$ and initial electron energy $\epsilon_i$ in a color 
logarithmic scale.
It is fully converged result, i.~e.\ it is independent of the method used to calculate it. 
Interestingly enough, the spectrum is qualitatively quite similar to the 2D spectrum for the CO$_2$
molecule \cite{CC1995,the_letter}.
The bulk of the spectrum is located at energies of the incident electron between 
2-4 eV. This is a consequence of the shape of the anion potential manifold 
(see Figs.~\ref{fig:PotentECD} and \ref{fig:Pot3DECD86}) and its location relative 
to the potential of the neutral molecule. 
The understanding of the detailed shape  is not trivial. 
 For small electron energy losses, the spectrum is discretized by vibrational 
frequencies whose ratio is approximately 3:1. But since this ratio is not exact,
the spectrum becomes quasi-continuous for energies above 1~eV. 
At the same time, we see 
that there is some selection mechanism that singles out narrower structures 
close to the diagonal threshold line. There are also diagonal rays appearing 
in the structure of the spectrum (better apparent in the decomposition of the 
spectrum according to symmetries). Both of these features were present in the case 
of CO$_2$, where we performed the detailed analysis \cite{partII}.

\begin{figure}[th]
\includegraphics[width=0.49\textwidth]{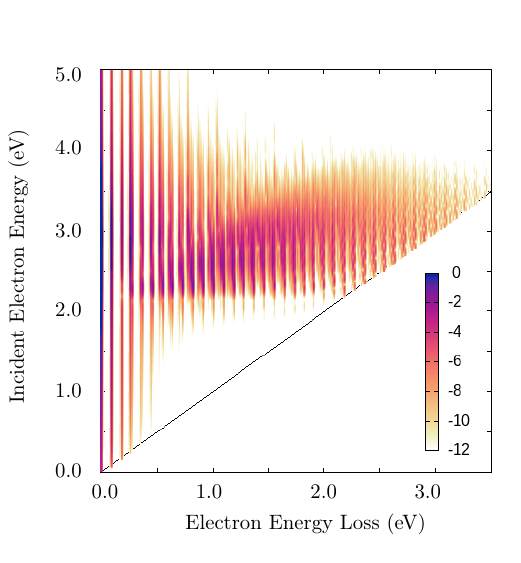}
\includegraphics[width=0.49\textwidth]{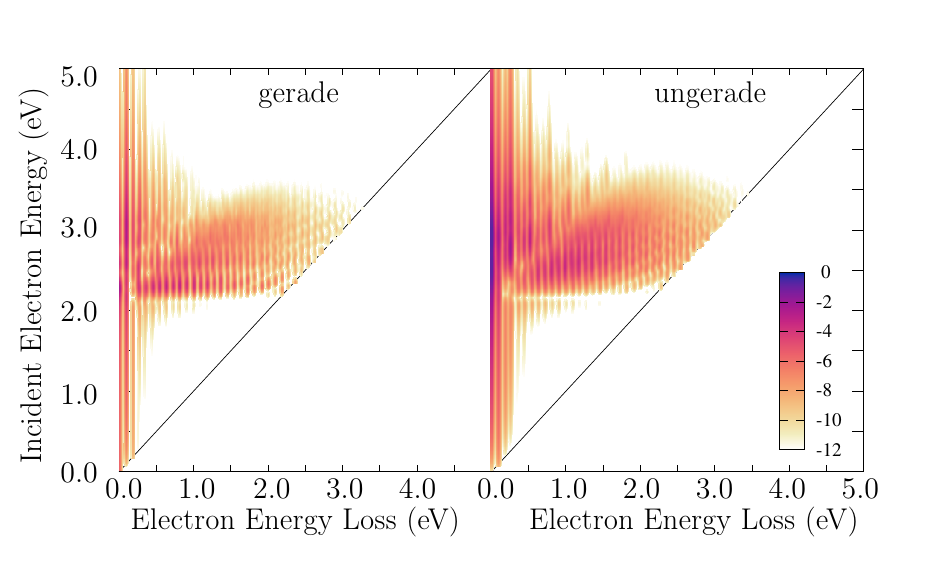}
\caption{\label{fig:EELS_B} 
           The same data as in Fig.~\ref{fig:EELS_A} but for the new model.
          }
\end{figure}

\subsection{2D spectrum for new model}

We proposed the new model above to consistently introduce the vibronic coupling in the 
level-shift operator in the ECD86 model and to test the iteration 
schemes to solve the dynamics in this model. The choice of the model parameters 
was guided by our experience with the diatomic molecules, but apart from 
that the choice is completely random. To our surprise, the resulting spectrum 
has a quite interesting intricate structure, which is furthermore similar 
to experimental data for some molecules, 
like benzene and its derivatives \cite{UNP_Allan}.
Particularly, we are speaking about the wedge-shaped structure with the tip 
touching the vertical axis at the incident electron energy of 2.2~eV, limited by this energy 
from below and limited approximately with the diagonal line corresponding to the electron 
energy loss $\Delta\epsilon=\epsilon_{i}+2$~eV from above. The origin of this structure is not clear
and since it is quite common in experimental data, we will dedicate the future study to this 
phenomenon. It indicates some selection mechanism in the dynamics that forces 
the system to skip through a region with small energy losses to large losses. 

\section{Conclusions}

We derived a generalization of the model of conical intersection in electronic continuum 
proposed originally by Estrada et al.~\cite{ECD1986} by including terms linear in the 
vibrational coordinates also in the term that couples the two discrete states of the original 
model to two partial waves of the electronic continuum. The generalization thus produces
quadratic terms in the nonlocal level-shift operator $F(E)$ that describes the dynamics of 
the vibrational excitation of the molecule by collision with an electron. 

We also implemented two Krylov-subspace iteration methods GMRES and COCG for solving 
the dynamics and calculation of 2D electron energy-loss spectra, and we tested the 
two methods on the original ECD86 model and our generalization. The Krylov-subspace 
methods are ideally suited for this kind of models because the multiplication 
by model functions expanded in polynomials of vibrational coordinates can be implemented 
very efficiently in the oscillator basis. For both models we observed a very good convergence 
of both methods even without preconditioning. The slower convergence of the COCG method is 
compensated by the simplicity of its implementation. The computational demands of GMRES also 
grow in the course of iteration procedure. The preconditioning by block-diagonal matrix 
works well only for some choice of the blocks (depends on the ordering of the basis). 

Out of all tested preconditioners, the $M^g$ preconditioner proved to be the most efficient
for both the two-dimensional models and our earlier realistic model of the CO$_2$
molecule. This seems to be a natural results since we exactly invert the blocks that involve
the discrete-state space and the coupling mode. On the other hand, the most time-consuming
preconditioner $M^d$ works rather badly taken into account that we invert the blocks
corresponding to the full vibrational space.
Thus, the $M^g$ preconditioner is the preconditioner of choice for more complicated models,
as the one for CO$_2$, where the unpreconditioned iterations are expected not to converge
for all energies of interest.


We believe that the methods tested here can be used for more complicated molecules 
to get better understanding of the 2D electron energy-loss spectroscopy. We plan a more 
extensive parameter study to obtain a deeper understanding of the results. The proposed
method is conceptually simple and can further be generalized in a straightforward way 
to include more anion states, more vibrational degrees of freedom and higher order 
polynomial functions. More challenging generalization will be needed to include also 
dissociative channels and anharmonicity in the neutral molecule.

\begin{acknowledgments}
 We gratefully acknowledge the financial support provided by the Czech Science Foundation
    Project No. 19-20524S and by the Charles University Grant Agency, Project No. 552120.  
\end{acknowledgments}

\bibliography{Refs}

\end{document}